\documentclass{article}

\usepackage{multirow}
\usepackage{hyperref}
\usepackage{graphicx}
\usepackage{amssymb, amsmath}
\begin{document}

\begin{center}

{\bf A comparison of two magnetic ultra-cold neutron trapping concepts using a Halbach-octupole array}
\vspace{5mm}

K.~LEUNG$^{*}$\footnote[2]{Currently at North Carolina State University, 2401 Stinson Dr.,
Riddick 421, Raleigh, NC 27695, USA.}, S.~IVANOV, F.~MARTIN, F.~ROSENAU, M.~SIMSON, AND O.~ZIMMER$^{*}$
\vspace{5mm}

Institut Laue-Langevin, 6 rue Jules Horowitz, 38042 Grenoble Cedex 9, France\\
$^*$E-mails: zimmer@ill.fr (OZ) and kkleung@ncsu.edu (KL)\\
\end{center}

\begin{abstract}
Preprint of an article published in Next Generation Experiments to Measure the Neutron Lifetime, pp. 145-154, \url{doi: 10.1142/9789814571678_0015$} $\copyright$ [copyright World Scientific Publishing Company] [\url{http://www.worldscientific.com/doi/abs/10.1142/9789814571678_0015}]
\\\\
This paper describes a new magnetic trap for ultra-cold neutrons (UCNs) made from a 1.2$\,$m long Halbach-octupole array of permanent magnets with an inner bore radius of $47\,{\rm mm}$ combined with an assembly of superconducting end coils and bias field solenoid. The use of the trap in a vertical, magneto-gravitational and a horizontal setup are compared in terms of the effective volume and ability to control key systematic effects that need to be addressed in high precision neutron lifetime measurements.
\end{abstract}

\section{Introduction}

The free neutron undergoes $\beta$-decay via ${\rm n} \rightarrow {\rm p} + {\rm e}^{-} + \bar{\rm \nu}_e$. Precise measurements of the mean lifetime $\tau_{\rm n}$ are used for obtaining the universal weak coupling constants of the nucleon from which one derives important semi-leptonic weak cross sections. They are also needed for searches of beyond Standard Model physics, and for calculations of primordial helium abundance in Big Bang Nucleosynthesis. These motivations are described in more details elsewhere in these proceedings, as well as in various review papers on the neutron particle physics field\cite{Abele2008a,Dubbers2011} or on $\tau_{\rm n}$ specifically \cite{Paul2009,Wietfeldt2011}.

Ultra-cold neutrons (UCNs) are free neutrons with kinetic energies less than the neutron optical potential of well-chosen materials so that they can be confined in a material ``bottle'' via total internal reflections. For instance, beryllium, a commonly-used material for UCN reflection, has $V_{\rm Be} = 252{\rm \, neV}$, corresponding to a velocity of $\approx 7{\rm \, m\,s^{-1}}$. UCNs stored in bottles\footnote{The exceptions being in-beam measurements\cite{Arimoto2012a,Nico2005a}.} for measuring $\tau_{\rm n}$ have been used in the most precise experiments to date\cite{Mampe1989a,Mampe1993a,Pichlmaier2010a, Arzumanov2012a, Serebrov2005, Serebrov2008}.

Due to the size of the neutron's magnetic moment $\mu_{\rm n}$ of $60{\rm \, neV\,T^{-1}}$, the Stern-Gerlach force can be used to confine UCNs also. This technique\cite{Vladimirskii1961,Paul1989,Huffman2000a} offers a method for confining UCNs free from energy-dependent wall losses that require corrections. While the traditional \emph{``counting the survivors''} scheme can be used for determining $\tau_{\rm n}$, extracting and detecting the charged decay products is also possible with magnetic traps. A strong motivation for using magnetic trapping is to improve upon material bottle experiments with a better control and reduction of systematic corrections\cite{Zimmer2000,Materne2009a,Walstrom2009,Ezhov2009}.

In magnetic bottles losses of UCNs due to depolarization at the weak field regions has been shown theoretically to be suppressible if a sufficiently strong bias field is used\cite{Steyerl2012a,Walstrom2009}. However, the effects caused by phase-space evolution of the UCN gas and the issue of how to effectively remove above-threshold UCNs must be carefully addressed by high-precision measurements. There is also the possibility of a gradual warming of the UCN spectrum, caused by magnetic noise or mechanical vibrations, which has largely been unexplored. With these issues in mind, two concepts using an Halbach-octupole array combined with a superconducting coil assembly are discussed in this paper.

Previously, the idea of using a UCN production volume integrated inside a horizontal, sliding magnetic bottle was presented\cite{Leung2009}. Its goal was to extract produced UCNs to vacuum in order to avoid losses due to interactions with the superfluid helium converter and also to avoid the dilution of the high density of UCNs offered by super-thermal production\cite{Golub1975}. In this scheme, due to the geometry of available cold neutron beams, the long axis of the trap is required to be horizontal. However, after further exploration of the idea, and combined with the recent success of a high-density superfluid helium UCN source using a vertical, window-less extraction system where transmission losses through windows are eliminated\cite{Zimmer2011}, we decided to employ a more traditional scheme.

\section{Design of magnetic fields}

\begin{figure}
\begin{center}
\includegraphics[width=4.5in]{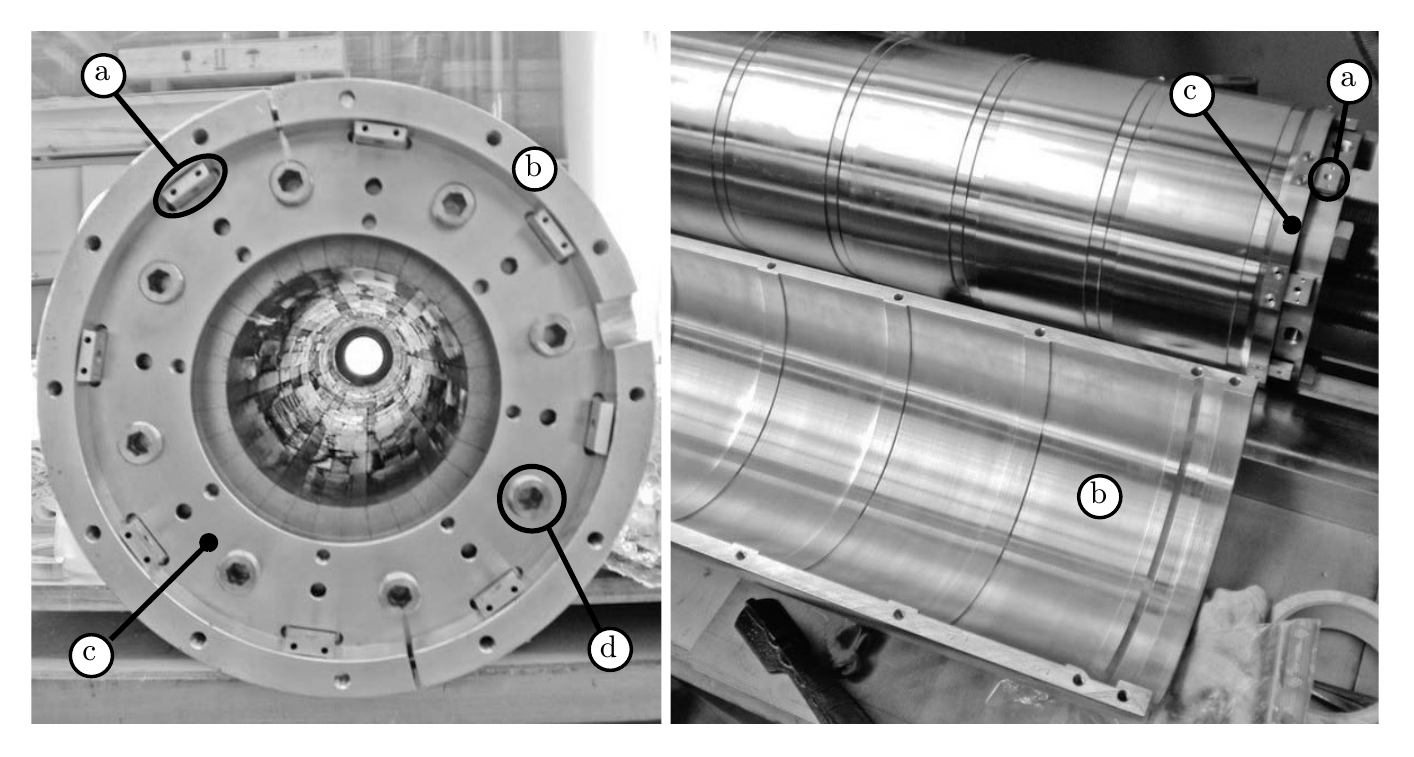}
\end{center}
\caption{The octupole magnet array assembly: (a) rods used to fix the rotation of the 12 modules, (b) split aluminum shell for holding the modules together, (c) stainless steel end plates fixed to the aluminum shell, and (d) brass pressing screws. The ribbed structure of the aluminum shell provides good thermal contact for cooling the magnets to $\sim\!120\,{\rm K}$. An individual module is shown in Ref.~\cite{Leung2009}.}
\label{fig:octupoleArrayPhoto}
\end{figure}

\begin{figure}
\begin{center}
\includegraphics[width=4.5in]{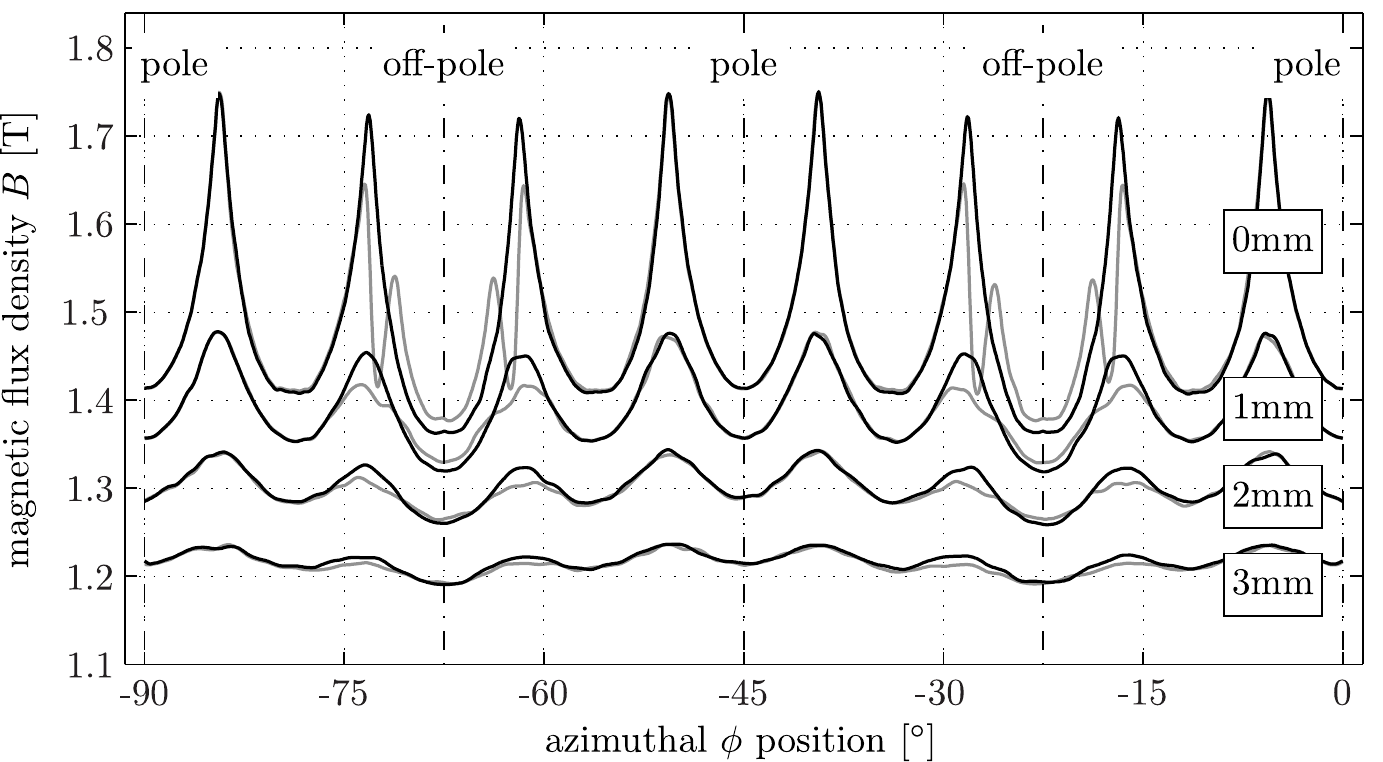}
\end{center}
\caption{2D finite element calculations of the Halbach-octupole. $B$ is plotted for $\phi$ spanning a quadrant in the $xy$-plane and for different distances from $R$ (indicated by the boxed numbers). The $\phi$ positions of the centers of the pole and off-pole pieces are indicated. The fainter lines for each wall distance are from modeling the demagnetized magnets with rounded corners. }
\label{fig:octupolefield}
\end{figure}

The central component of the discussed setups is the $1.2{\rm \, m}$ long 32-piece Halbach-type\cite{Halbach1980} octupole array for radial UCN confinement (shown in Fig.~\ref{fig:octupoleArrayPhoto}). The octupole has an inner bore radius $R = 47\,{\rm mm}$ and a nominal $B = 1.3{\, {\rm T}}$ at its surface (at room temperature). While the field near the center is that of an ideal octupole $B(\rho) \propto \rho^3$, deviations appear near $R$ due to the discrete number of magnets. The flux density $B$ from 2D finite element calculations with FEMM \cite{Meeker2006a} at different azimuthal angles $\phi$ and distances from $R$ is shown in Fig.~\ref{fig:octupolefield}. The weakest $B$ for a fixed radial position $\rho$ occurs at the off-pole pieces\footnote{The pieces are called off-pole when they have their magnetization vector parallel or anti-parallel to the $\hat{\phi}$ unit vector at the centre of that piece.}. 

These calculations also show that the magnetic material at the off-pole pieces' inner corners are the most susceptible to demagnetization. To model how this might affect $B$ in the bore, these corners were rounded out with a $1.5\,{\rm mm}$ radius (and replaced with an air gap). This causes sharp dips right at the corners and a slight enhancement at the center of the piece (also in Fig.~\ref{fig:octupolefield}). These features become insignificant at distances $>\sim 2\,{\rm mm}$ from $R$. Conclusions are the same for different corner geometries (quarter-circular and chamfer cut). Besides from demagnetization, if chips in the brittle NdFeB material exist, then similar dips could exist. It is therefore wise to keep UCNs from exploring too close to these regions.

\begin{figure}
\begin{center}
\includegraphics[width=4.5in]{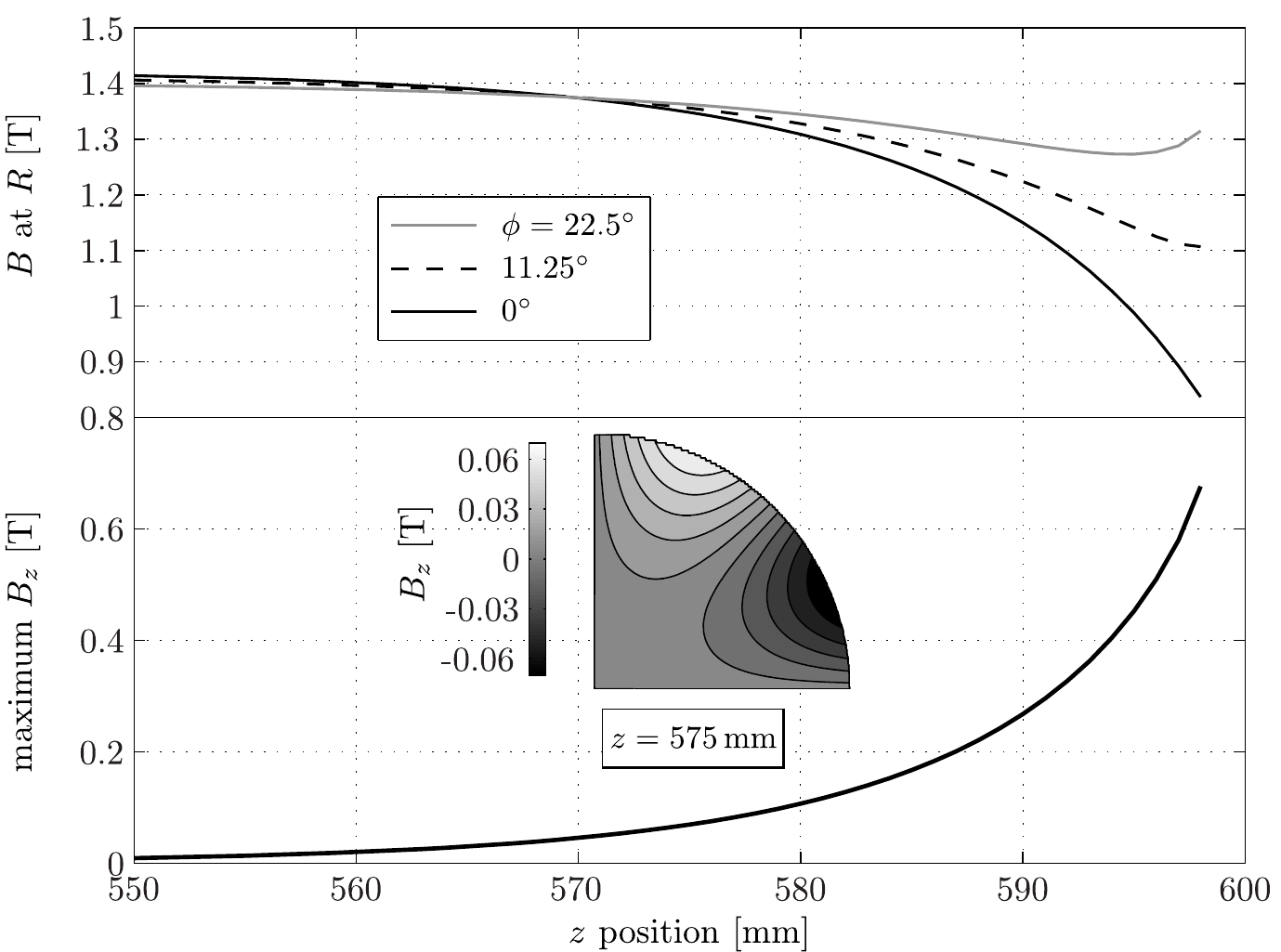}
\end{center}
\caption{The field at the ends from solely the octupole array: $B$ at $R$ becomes weaker and strong $B_z$ components start to emerge. A quadrant slice of $B_z$ at $z=575\,{\rm mm}$, the central position of the end coil placed closest to the end of the array (see Fig.~\ref{fig:superconductingcoilfields}). }
\label{fig:octupoleArrayEndFields}
\end{figure}

To understand the field at the ends of the array 3D finite element calculations were performed with RADIA\cite{Chubar1998a}. This revealed that $B(R)$ drops to $<0.8\,{\rm T}$ at the end of the array and there exist axial components $B_z \approx 0.8\,{\rm T}$ that are strongest near the surface of the pole pieces. These features are shown in Fig.~\ref{fig:octupoleArrayEndFields} and have been confirmed with Hall probe measurements\cite{Fraval2009}. 

\begin{figure}
\begin{center}
\includegraphics[width=4.5in]{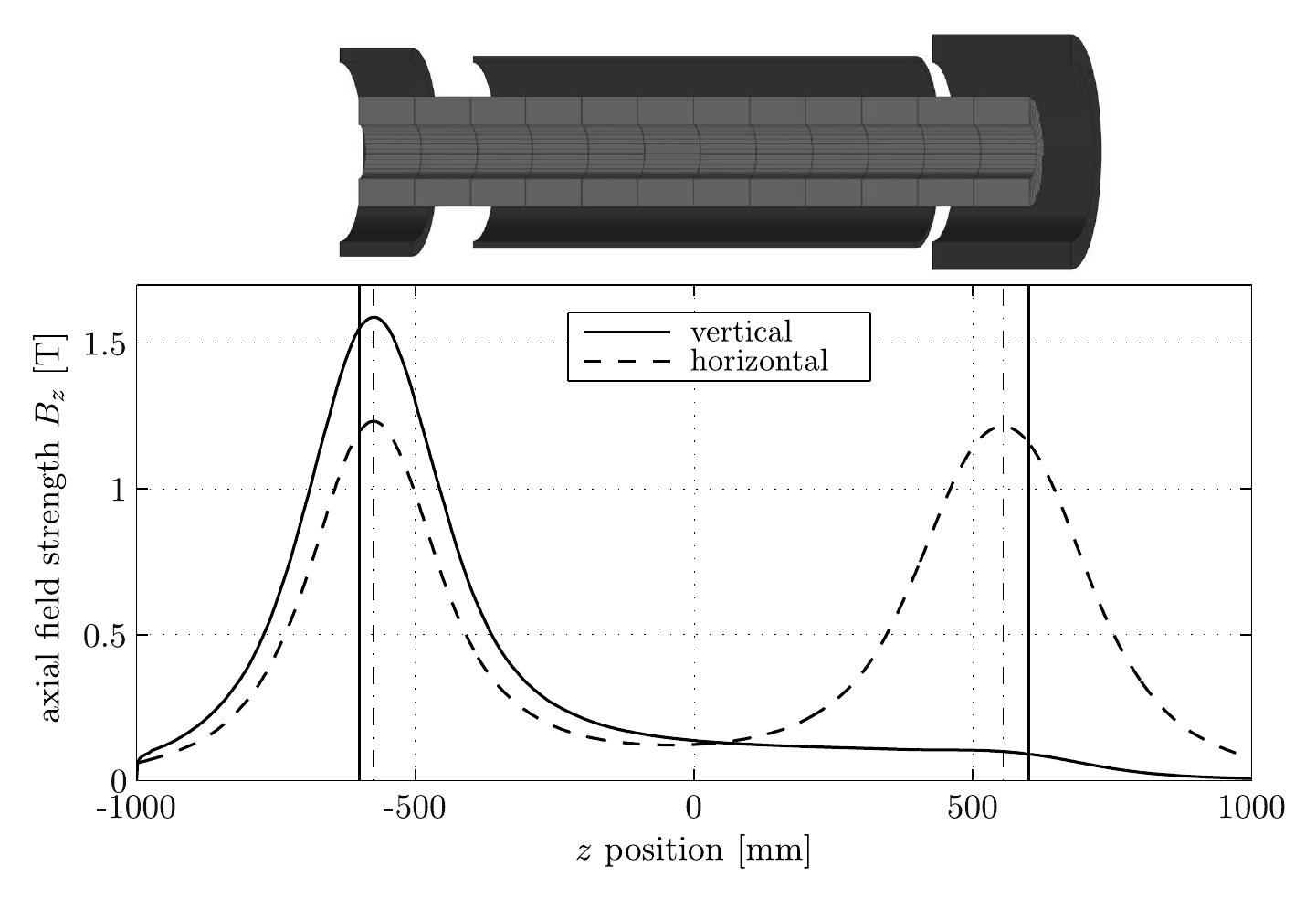}
\end{center}
\caption{\emph{Top:} a to-scale drawing of the octupole array and the surrounding assembly of superconducting coils. \emph{Bottom:} the axial field from the coils for the two trapping configurations. The vertical solid lines show the end of the array and dash-dotted lines the center of the end coils.}
\label{fig:superconductingcoilfields}
\end{figure}

A superconducting coil assembly (depicted in Fig.~\ref{fig:superconductingcoilfields}) consisting of a small end coil, a bias field solenoid, and a large end coil is used for axial UCN confinement and removal of the low-field region. The fields that can be produced by these coils when they are run at their maximum current of $300\,{\rm A}$ are $1.7\,{\rm T}$, $1.2\,{\rm T}$, and $5\,{\rm T}$\footnote{This allows focussing decay protons onto a small detector. This might be done in later experiments but will not discussed in this paper.}, respectively. The $30\,{\rm cm}$ inner diameter of the coils produces only small $B_\rho$ components at $R$, and hence there is only a small cancellation with $B_\rho$ from the octupole array. The end coils are slotted inside the octupole array so that the region of strong $B_z$ cancellation is situated away from the trapped UCNs. Two magnetic field configurations from the coils that will be discussed in the next section are also shown in Fig.~\ref{fig:superconductingcoilfields}. The bias field is chosen to be $>0.1\,{\rm T}$ for preventing depolarization in the low field region. 

Combining the 3D calculations with all field sources, it was shown that demagnetization of the permanent magnets can be avoided by cooling them to $\sim\!120\,{\rm K}$, which leads to a strong increase in the coercivity. This cooling also increases $B$ at $R$ by $\sim\!10\%$, which is not included in the discussion to keep the estimates of the trap depth conservative.

\section{Trapping potential and effective volume}

\begin{figure}
\begin{center}
\includegraphics[width=4.5in]{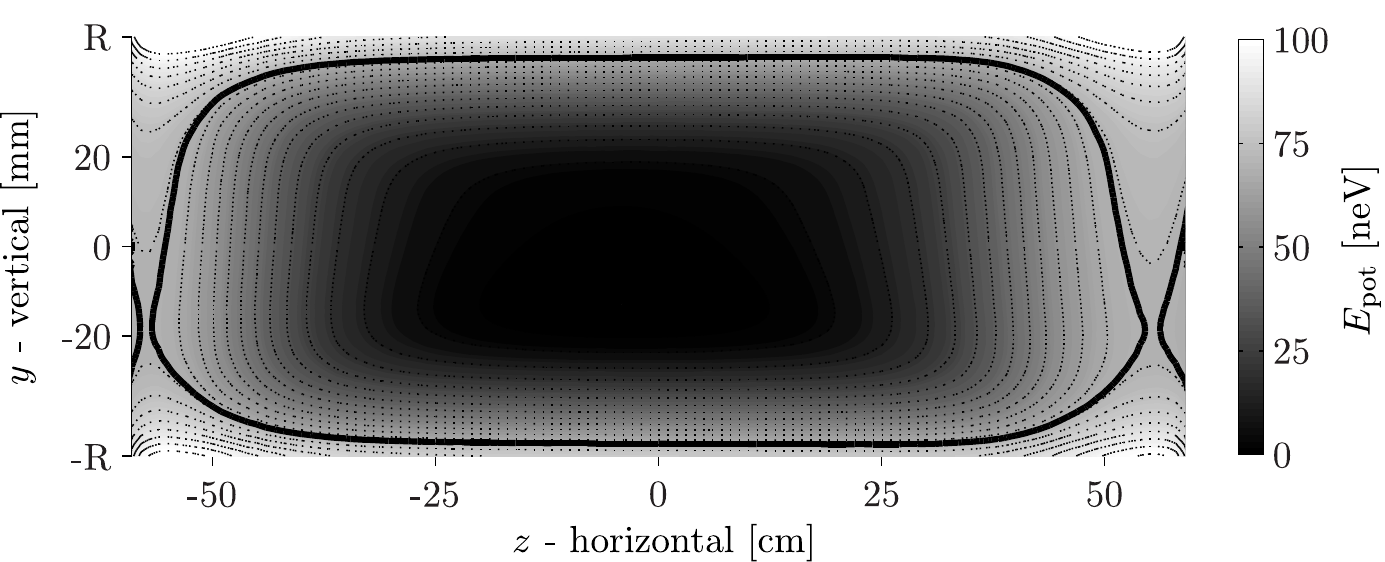}
\end{center}
\caption{A map of $E_{\rm pot}$ for the horizontal configuration with the axial fields shown in Fig.~\ref{fig:superconductingcoilfields} and for the $\phi=90^{\circ}$ slice aligned vertically. The thick line shows the contour for $E_{\rm trap} = 63\,{\rm neV}$ and the dotted contour lines are placed at increments of $5\,{\rm neV}$.}
\label{fig:horizontalEpotContour}
\end{figure}

\begin{figure}
\begin{center}
\includegraphics[width=4.5in]{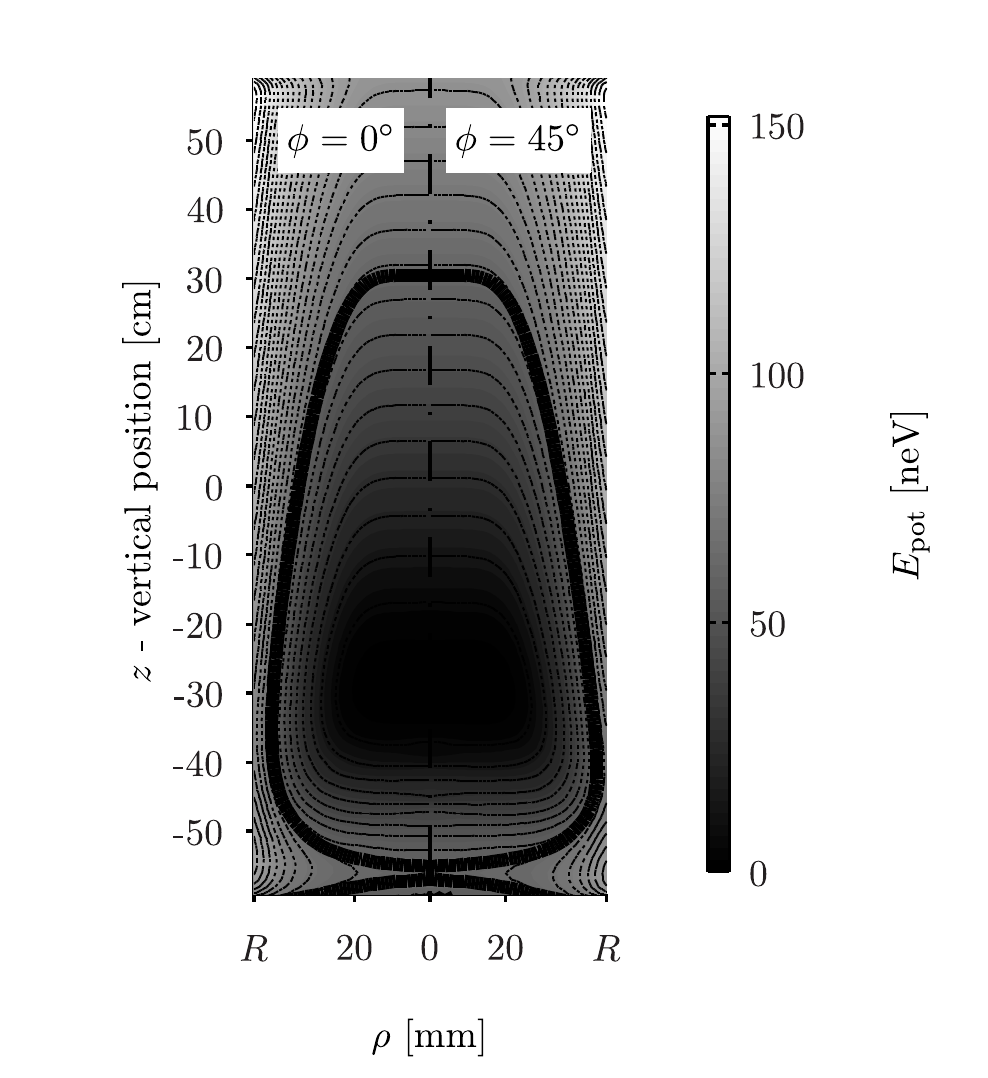}
\end{center}
\caption{A map of $E_{\rm pot}$ for the vertical configuration with the axial fields shown in Fig.~\ref{fig:superconductingcoilfields} for $\phi=0^{\circ}$ (maximum $B_\rho$ reinforcement) and $\phi=45^{\circ}$ (maximum $B_\rho$ cancellation) slices. The thick solid line is for $E_{\rm trap} = 48\,{\rm neV}$ and the dotted contour lines are placed at increments of $5\,{\rm neV}$.}
\label{fig:verticalEpotContour}
\end{figure}

The total energy of a neutron  $E_{\rm n}$ at a point $\vec{r} = (x,y,z)$ in space is given by:
\begin{multline}
E_{\rm n} = E_{\rm kin}(\vec{r}) + E_{\rm pot}(\vec{r}) = E_{\rm kin}(\vec{r}) + V_{\rm grav}(z) + V_{\rm mag}(\vec{r}) \\= E_{\rm kin}(\vec{r}) + mgz \pm \mu_{\rm n} B(\vec{r}) \;,
\end{multline}
where $E_{\rm kin}$ is the kinetic energy, $m$ is the neutron mass, $g$ is the gravitational acceleration, and $z$ is the height. A constant offset is neglected to place $E_{\rm pot}$ at the potential minimum of the trap. In the last term, the plus sign refers to the \emph{low field seeking} spin-state. Maps of $E_{\rm pot}$ for a horizontal configuration, where both end coils are required for axial confinement, and for a vertical configuration, where only the bottom end coil is required due to the gravitational potential $mg = 102\,{\rm neV\,m^{-1}}$  (\emph{magneto-gravitational} trap), are shown in Figs.~\ref{fig:horizontalEpotContour} and \ref{fig:verticalEpotContour}. In order to populate the trap with UCNs, current in the small end coil will be lowered temporarily to reduce $E_{\rm pot}$.

The depth of a pure magnetic trap $E_{\rm trap}$ is defined by the maximum $E_{\rm n}$ a UCN can have without the possibility of it exiting the trap at the ends or making contact with material walls. Since an inner tube will be placed inside the magnet bore, a reduction in the radius of the trap of $1.5\,{\rm mm}$ is taken into account for determining $E_{\rm trap}$. 

\begin{figure}
\begin{center}
\includegraphics[width=4.5in]{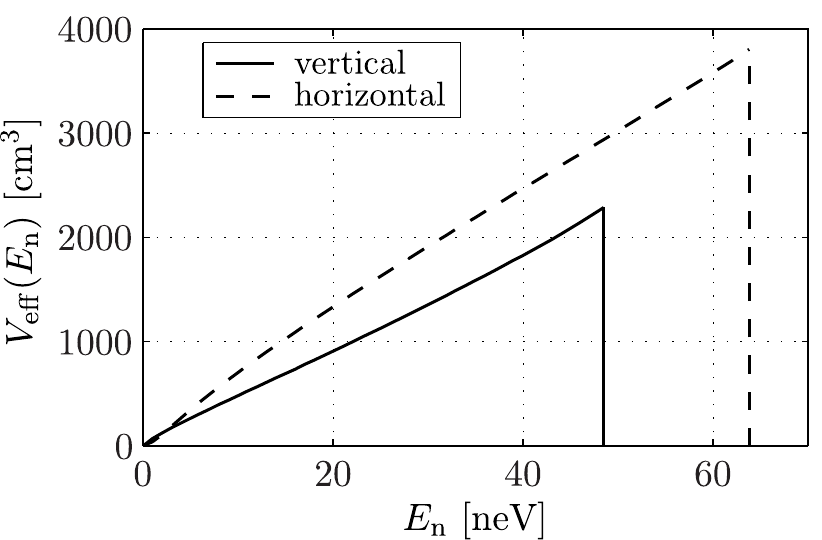}
\end{center}
\caption{The UCN energy-dependent effective volumes for the two configurations.}
\label{fig:effectiveVolumes}
\end{figure}

A UCN clearly cannot explore regions in a trap where $E_{\rm n}<E_{\rm pot}(\vec{r})$ so that the accessible volume is energy-dependent. Thus, it is useful to define the concept of the effective volume\cite{Golub1991} $V_{\rm eff}(E_{\rm n})$, which can be expressed as:
\begin{equation}
\label{eq:HOPEVeffEucn}
V_{\rm eff}(E_{\rm n}) =  {\rm Re}\left[ \int_{V} \,\sqrt{\frac{ E_{\rm n} - E_{\rm pot}(\vec{r})}{E_{\rm n}}}\,{\rm d}V\right] \:,
\end{equation}
where $\int_V$ is over volume of the trap and ${\rm d}V$ is the volume element at $\vec{r}$. This is shown for the two configurations in Fig.~\ref{fig:effectiveVolumes}. The total number of UCNs stored in the bottle is then given by: $\int_0^{E_{\rm trap}} n(E_{\rm n})\,V_{\rm eff}(E_{\rm n}) \,{\rm d}E_{\rm n}$, where $n(E_{\rm n})$ is the energy-dependent spectral UCN density. If this is calculated for the typical Maxwellian  spectrum, where $n(E_{\rm n}) \propto \sqrt{E_{\rm n}}$, then the number that can be stored in the horizontal configuration is $\sim \! 2.5$ times greater than in the vertical.

\section{Advantages of the vertical configuration}
Despite the larger number of UCNs that can be stored in the horizontal setup, there are distinct advantages for the vertical setup in terms of systematic effects.

In the horizontal configuration, the two magnetic mirrors can trap the charged decay products. \emph{In-situ} counting with a high efficiency is possible only by extracting the protons with high-voltage electrodes ($> \!5\,{\rm kV}$) due to their low kinetic energy of up to $750\,{\rm eV}$ only. The trapped electrons can produce ions and electrons in the residual gas of the vacuum, and the presence of the electric field can accelerate them causing production of secondary charges up to electrical break down. For instance, even in the $a$SPECT neutron $\beta$-decay apparatus, with its excellent vacuum conditions, this effect has been problematic\cite{Zimmer2000a,Konrad2009a}. The vertical configuration avoids the electron trapping problem since the upper coil is not required for UCN confinement. Moreover, this allows electron counting as the detection scheme and thus removes the need for the high-voltage system.

A key systematic effect is poor cleaning of above-threshold UCNs from the trap. For a horizontal configuration, a scheme of ramping up and down the bias field has been used for this, however this results in a significant loss ($\sim 50\%$) of UCNs\cite{Dzhosyuk2005}. The use of a UCN reflecting paddle along the length of a trap to induce mode-mixing reflections has also been suggested\cite{Bowman2005}. However, removing the paddle from the reach of UCNs after cleaning by rotation or by sliding it out of the trap causes undesired doppler heating of the UCNs (for the latter case, heating can occur after non-specular reflections\cite{Leung2013a}). In the vertical, magneto-gravitational configuration, UCNs have to fall and reflect off the bottom of the trap. The insertion of a piston that induces mode-mixing reflections from the bottom can thus be used for cleaning. Furthermore, retracting it does not cause doppler heating. The details of this procedure will be published elsewhere.

Finally, an advantage offered by this magnetic bottle is the large end opening area to volume ratio. When it is vertical, the opening at the bottom will allow the live monitoring of depolarized or warmed UCNs with a high efficiency. This will allow direct measurements of these key systematics. Furthermore, if a fill-and-empty scheme is used to determine $\tau_{\rm n}$---as will be the case of initial measurements---emptying of the trap from the bottom will have weak sensitivity to the phase-space evolution of the UCNs, a key topic discussed in these proceedings.

\section{Conclusion}

A detailed description of the magnetic fields from the Halbach-octupole array and the superconducting coil assembly design has been given. The trapping potential and the effective volumes for a horizontal and a vertical geometry were compared, with the former allowing the storage of a factor $\sim\!2.5$ more UCNs. However, the advantages for controlling key systematic effects offered by the vertical setup outweighs the decreased statistics, which is addressed with the development of the new, compact superfluid helium UCN source\cite{Zimmer2011}. This has led to the magneto-gravitational design for our planned high-precision neutron lifetime measurements.

\bibliographystyle{ws-procs9x6}
\bibliography{Leung_neutronworkshop}

\end{document}